\documentclass[aps,physrev,reprint,superscriptaddress,floatfix]{revtex4-2}
\usepackage[dvipsnames]{xcolor}
\usepackage{amsmath,amssymb,graphicx,bm,microtype} 
\usepackage[colorlinks,allcolors=blue!50!black]{hyperref}
\usepackage[capitalise]{cleveref}
\usepackage[all]{hypcap} 

\newcommand{\USyd}{School of Chemistry, University of Sydney, Sydney,  UNSW 2006, Australia}

\newcommand{\UQ}{School of Mathematics and Physics, The University of Queensland, 4072, Australia}

\begin{document}

\title{
 Fingerprints of collective magnetic excitations in inelastic electron tunneling spectroscopy
}

\author{F. Rist}
\affiliation{\UQ}
\author{H. L. Nourse}
\affiliation{\USyd}
\author{B. J. Powell}
\email{powell@physics.uq.edu.au}
\affiliation{\UQ}

\date{\today}

\begin{abstract}
Spin-flip inelastic electron tunneling spectroscopy allows magnetic materials to be probed at the single-atom level via scanning tunneling microscopy. Previously, the local spectral weight of spin excitations of small systems has been deduced from discrete steps in the differential conductance. However, this is not viable for large systems. We show that the local spin density of states can be measured via the double differential conductance. This contrasts with elastic measurements where the local density of electronic states is deduced from the differential conductance. We study the tunneling currents of the spin-1/2 and -1 Heisenberg chains and propose a method to probe zero-frequency modes.
\end{abstract}
\maketitle





Spin-flip inelastic electron tunneling spectroscopy (IETS) has emerged as a powerful probe of magnetic excitations at the atomic scale, particularly in low-dimensional systems~\cite{cyrus_single_atom_STM_IETS,heinrich_single_atom, heinrich_spin_chain,Haldane_nanographene}. When combined with a scanning tunneling microscope (STM), IETS allows local probes of the magnetic modes of adatoms through discrete steps in the differential conductance ($dI/dV$). The voltage of these steps reflects the threshold energy required to induce a spin excitation, while the step height corresponds to the local spin spectral weight of the excitation \cite{Loth_2010, Cyrus_anisotropy_2007,Ternes_2015,cyrus_single_atom_STM_IETS}. This has allowed IETS to be a key experimental technique for studying low-dimensional magnetic systems, as it allows \textit{local} probes of collective spin-wave excitations~\cite{spin_waves_imaging}, quantum spin liquids~\cite{QSL_IETS_theory,QSL_IETS_theory_edge}, and topological edge states in spin chains~\cite{Mishra_nanographene,Localprobes_heisenberg}.

 Recently, there has been considerable interest in self-assembled and artificially constructed spin chains on surfaces for studies of Majorana edge modes and collective spin excitations~\cite{Kamlapure_large_fe_chains_YSR,Xiaoshuai_Heisenberg_chain_synthesis,Xuelei_Heisenberg_chain_spin_half_surface,Zhao_spinon_2025}. However, for large spin systems, identifying the spin spectral features becomes challenging. Relying on distinct, discrete steps in the differential conductance 
 to distinguish the different spectral weights is difficult as discrete features merge to form a continuum in large systems. Understanding how to interpret IETS differential conductance features in the absence of distinct conductance steps in large magnetic systems remains an open question.

In this Letter, we show that the local spin spectral function, \cref{eq:main_result}, and spin local density of states (LDOS), \cref{eq:DOS_relation}, of a magnetic system can be obtained from the double differential conductance ($d^2I/dV^2$) for any system size. By relating the double-differential conductance to the LDOS, we can identify the features in the (double) differential conductance features that correspond to collective magnetic excitations and zero-frequency modes in the LDOS.

 To illustrate this, we show how the double differential conductance can be used to identify spinons and magnons in the spin-1/2 and spin-1 antiferromagnetic Heisenberg chains in the thermodynamic limit. We also show that zero-frequency modes of the spin LDOS, which are typically challenging to identify unambiguously, can be measured through local spectral sum rules for the LDOS. The characterization of zero-frequency modes is crucial for identifying Majorana modes, which may play an important role in topological quantum computing \cite{Stephan_review_2025,Majorana_zero_modes}.

\textit{IETS.}  The inelastic tunneling current in IETS can be modeled using an exchange Hamiltonian that couples the STM tip and the metallic substrate with the spin of the probed adatom~\cite{theory_STM_rossier,Fransson_2010,Applebaum_1967,Anderson_kondo_1966}. IETS measurements are undertaken for adatoms that are effectively decoupled from the substrate's electron bath, meaning that the tunneling between the adatom and the substrate can be treated perturbatively. It has previously been shown that the contribution of inelastic tunneling to the current for a non-magnetic tip is~\cite{theory_STM_rossier, Delgado_spin_dynamics_2010,Ternes_2015,Fransson_2010},
\begin{equation}
I_j=\frac{g_S}{e}  \int_{-\infty}^\infty\int_{-\infty}^\infty d\epsilon d\epsilon' \ \mathcal{S}_j(\epsilon-\epsilon',\beta)F(\epsilon,\epsilon',eV)\ ,\label{eq:Rossier_current} \end{equation}
where $e$ is the elementary charge, $g_S$ is a dimensionless parameter which is linearly dependent on the electronic density of states of the tip and substrate and the coupling between the probed spin and the STM tip and substrate \cite{theory_STM_rossier}, $V$ is the voltage difference between the STM tip and substrate, $\beta=1/k_BT$ is the inverse temperature, $\mathcal{S}_j(\omega,\beta)$ is the local spin spectral function for the adatom on site $j$, $F(\epsilon,\epsilon',eV)  = f(\epsilon)[1-f(\epsilon'+eV)]-f(\epsilon+eV)[1-f(\epsilon')]$,
and $f(\epsilon)$ is the Fermi-Dirac distribution. The inelastic tunneling current is therefore a probe of the local spin spectral function on site $j$, 
\begin{align}
\mathcal{S}_j(\omega,\beta)
&=\frac{1}{2\pi}\int_{-\infty}^\infty dt \ \chi_j(t,\beta)\ e^{i\omega t},
\label{eq:spectral_density}
\end{align}
where  the local dynamical spin correlation function at site $j$  is
\begin{equation}
\chi_j(t,\beta) =\chi'_j(t,\beta)+i\chi_j''(t,\beta) = \sum_{\alpha\in x,y,z}\mathrm{Tr}[e^{-\beta\hat{H}}\hat{S}^\alpha_j(t)\hat{S}_j^\alpha].\label{eq:Correlation_def}
\end{equation}
We make the standard assumption that the inelastic tunneling is only significant for the spin directly underneath the tip due to the exponential decay of the electron tunneling strength~\cite{Bardeen_1961,theory_STM_rossier,Rossier_delago_review, Localprobes_heisenberg,Mishra_nanographene}.

\textit{Spin LDOS from the double differential conductance.} 

The inelastic tunneling current, \cref{eq:Rossier_current}, can be written as
\begin{align}
I_j&=\frac{g_S}{2\pi e} \int_{-\infty}^\infty dt \  \chi_j(t,\beta)\tilde{f}(t,0) \bigg{(}\tilde{f}(t,-eV)- \tilde{f}(t,eV)\bigg{)},
\end{align}
where 
$\tilde{f}(t,\omega_0)
\equiv \int_{-\infty}^\infty d\omega f(\omega+\omega_0) e^{-i\omega t}
= \pi\delta(t) - i({e^{-i t\omega_0}}/{t})\sqrt{\mathcal{G}(t,\beta)}$ 
and
$\mathcal{G}(t,\beta) = {(\pi t/\beta)^2}/{\sinh^2(\pi t/\beta)}$
describes the thermal broadening. Hence, the inelastic tunneling current and its derivatives are
\begin{align}
I_{j}&={g_S} \bigg{[}V\chi_j(0,\beta)-\frac{i}{e\pi}\int_{-\infty}^\infty dt \ \chi''_j(t,\beta)\frac{e^{iteV}}{ t^2}\mathcal{G}(t,\beta)  \bigg{]},
\\
\frac{\partial I_{j}}{\partial V}&=g_S\bigg{[}\frac{1}{\pi}\int_{-\infty}^\infty dt \ \chi''_j(t,\beta)\frac{e^{iteV}}{ t}\mathcal{G}(t,\beta) 
+\chi_j(0,\beta) \bigg{]} ,\label{eq:diff_conductance}\\
\frac{\partial^2 I_{j}}{\partial V^2}&=\frac{ieg_S}{\pi} \int_{-\infty}^\infty dt\   \chi''_j(t,\beta)\mathcal{G}(t,\beta)e^{ieVt}. \label{eq:double_diff}
\end{align}
We can take the inverse Fourier transform of \cref{eq:double_diff} and, for $\omega>0$, use the quantum fluctuation-dissipation theorem \cite{Kubo_1966} to obtain
\begin{equation}
\begin{split}\mathcal{S}_j(\omega,\beta) & =\frac{1}{e g_S (1-e^{-\hbar\beta \omega})} \\& \phantom{=} \times\int_{-\infty}^\infty \int_{-\infty}^\infty dt\  dV\frac{ e^{-i(\omega-eV)t} }{\mathcal{G}(t,\beta)}\frac{\partial^2 I_{j}(V)}{\partial V^2}\label{eq:main_result} \ .
\end{split}
\end{equation}
The single-magnon LDOS is defined \cite{White_spectral_function}  as
\begin{align}
\mathcal{D}_j(\omega)&\equiv \sum_{M} \delta(\omega - E_M+E_0) |\langle \Phi_j|M \rangle|^2,\\
&=\frac{1}{2\pi}\sum_{\alpha=x,y,z}\sum_{N_0}\int_{-\infty}^\infty dt \ \langle N_0|\hat{S}^\alpha_j(t)\hat{S}^\alpha_j|N_0\rangle e^{i\omega t} \label{eq:spin_dos_def},
\end{align}
where $|N_0\rangle$ is a groundstate of the spin system and $|\Phi_j\rangle=\sum_{N_0}\sum_{\alpha\in x,y,z}\hat{S}_j^\alpha|N_0\rangle$ are the set of states that are connected to a groundstate by a local spin-flip. Hence, the spin LDOS can be calculated from the double differential conductance as
\begin{align}\mathcal{D}_j(\omega)
&=\lim_{\beta\rightarrow\infty}\frac{1}{ e g_S  (1-e^{-\hbar\beta \omega})}\frac{\partial^2 I_{j}(\omega)}{\partial V^2}\label{eq:DOS_relation} .
\end{align}


\begin{figure}
\begin{centering}
\includegraphics[width=\linewidth]{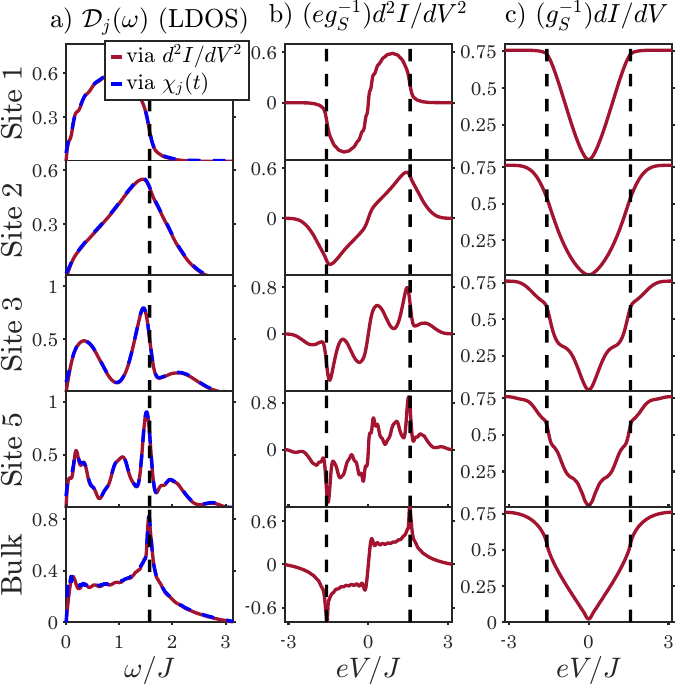}
\caption{\label{fig:spinhalf} 
Spin-1/2 Heisenberg chain in the thermodynamic limit. 
(a) Spin LDOS, $\mathcal{D}_j(\omega)$, calculated from the double differential conductance, $d^2I_j/dV^2$, [red line] matches the LDOS calculated from the local dynamical spin correlation function, $\chi_j(t,\beta=0)$ [blue line]. 
(b) In the bulk ($j=N/2$) the double differential conductance, $d^2I_j/dV^2$, is strongly peaked at the des Cloizeaux-Pearson spin-wave frequency ($eV=\pi J/2$; dashed black line). However, near the edge of the chain the peak in $d^2I_j/dV^2$ the peak broadens, and on the edge site ($j=1$) only a broad hump remains, centered at approximately half the des Cloizeaux-Pearson spin-wave frequency.
(c) Similarly,  the differential conductance, $dI_j/dV$, shows a kink at the des-Cloizeaux Pearson spin-wave frequency, which becomes less pronounced near the edge.
} 
\end{centering} 
\end{figure}

\begin{figure}
\begin{centering}
\includegraphics[width=\linewidth]{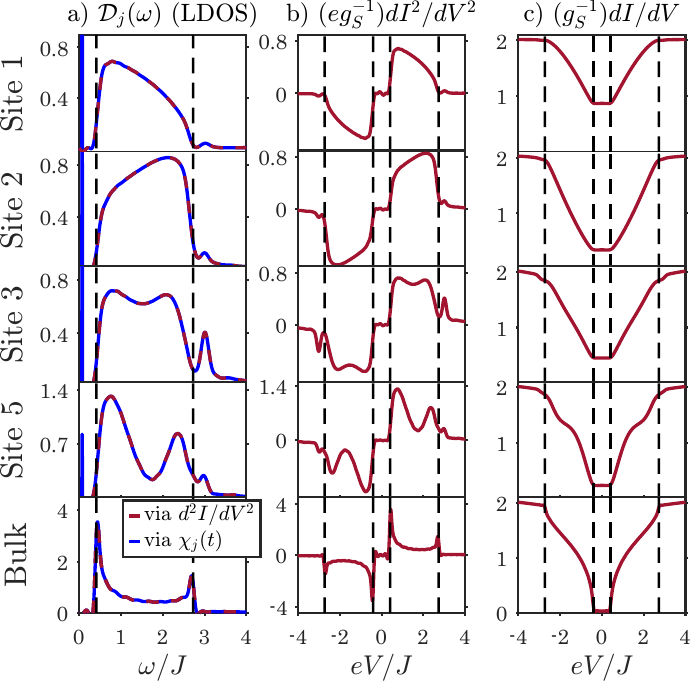}
\caption{\label{fig:spinone}%
Spin-1 Heisenberg chain in the thermodynamic limit. 
(a) Spin LDOS, $\mathcal{D}_j(\omega)$, calculated from the double differential conductance, $d^2I_j/dV^2$, [red line] matches the LDOS calculated from the local dynamical spin correlation function, $\chi_j(t,\beta=0)$ [blue line] via \cref{eq:spin_dos_def}, except at $\omega=0$. The zero-frequency LDOS is not captured due to the non-analyticity of the quantum-fluctuation-dissipation theorem at $\omega=0$. In particular,  the LDOS extracted from $d^2I_j/dV^2$ misses the zero frequency modes associated with the edge modes, see plots for small $j$. 
(b) In the bulk clear peaks are found in $d^2I_j/dV^2$ at the Haldane gap ($\omega \sim 0.41 J$; marked by dashed black lines) and the maximum in the single-magnon dispersion ($\omega \sim 2.725 J$ \cite{White_spectral_function}; marked by dashed black lines). As we move towards the edge of the chain these broaden and shift energies.
(c) Corresponding kinks are observed in $dI_j/dV$ at the Haldane gap and the single-magnon maximum. The zero-bias differential conductance increases as we move nearer the edge of the chain.
}
\end{centering}
\end{figure}

\textit{Numerical methods.} \label{sec:method} To test and explore the implications of \cref{eq:main_result} and \cref{eq:DOS_relation}, we calculate the LDOS, the spin spectral function, and the tunneling currents of the spin-$1/2$ and spin-$1$ antiferromagnetic Heisenberg chains,  
\begin{equation} 
\hat{H}=J\sum_{j=1}^{N} \hat{\textbf{S}}_j \cdotp \hat{\textbf{S}}_{j+1},\label{eq:heisenberg}
\end{equation} 
where $L$ is the length of the spin-chain and $J$ is the exchange coupling strength.

We calculated the tunneling currents using density matrix renormalization group (DMRG) in the matrix product state (MPS) formalism~\cite{itensor,thermal_DMRG_ancilla,white1992dmrg,ulrich1,white1993dmrg,DMRG_ostlund_1995}. The dynamical local spin correlators, $\chi_j(t)$, of the groundstate are calculated using time-evolving block decimation (TEBD) with a second-order Trotter decomposition~\cite{White_spectral_function}. These dynamical local spin correlators are then used to calculate the local spin spectral function and LDOS, Eqs. (\ref{eq:spectral_density}) and (\ref{eq:spin_dos_def}), the differential conductance, \cref{eq:diff_conductance}, and the double differential conductance, \cref{eq:double_diff}. We also calculate the finite-temperature tunneling currents for the spin-$1/2$ Heisenberg model. To do so, we use a maximally entangled ansatz, $|\psi\rangle =(1/\sqrt{2})^N\bigotimes_{j=1}^N\sum_{\sigma\in \uparrow,\downarrow}|\sigma \overline\sigma\rangle_j$,
where $\sigma$ represents the spin of the physical site and $\bar{\sigma}$ represents the ancilla spin. We then evolve the physical sites in imaginary time to the desired temperature using TEBD to obtain the thermal state~\cite{thermal_DMRG_ancilla}.

Time evolution using TEBD increases the entanglement of the MPS, reducing its accuracy for long time evolutions, especially in the case of finite-temperature tunneling currents~\cite{White_spectral_function}. To mitigate the error for long times, we use linear prediction~\cite{White_spectral_function} to extrapolate the local spin correlators to later times to minimize both spectral leakage and entanglement growth within the MPS. We employ a Hanning window after linear prediction to further minimize spectral leakage.

\begin{figure}
\begin{centering}
\includegraphics[width=\linewidth]{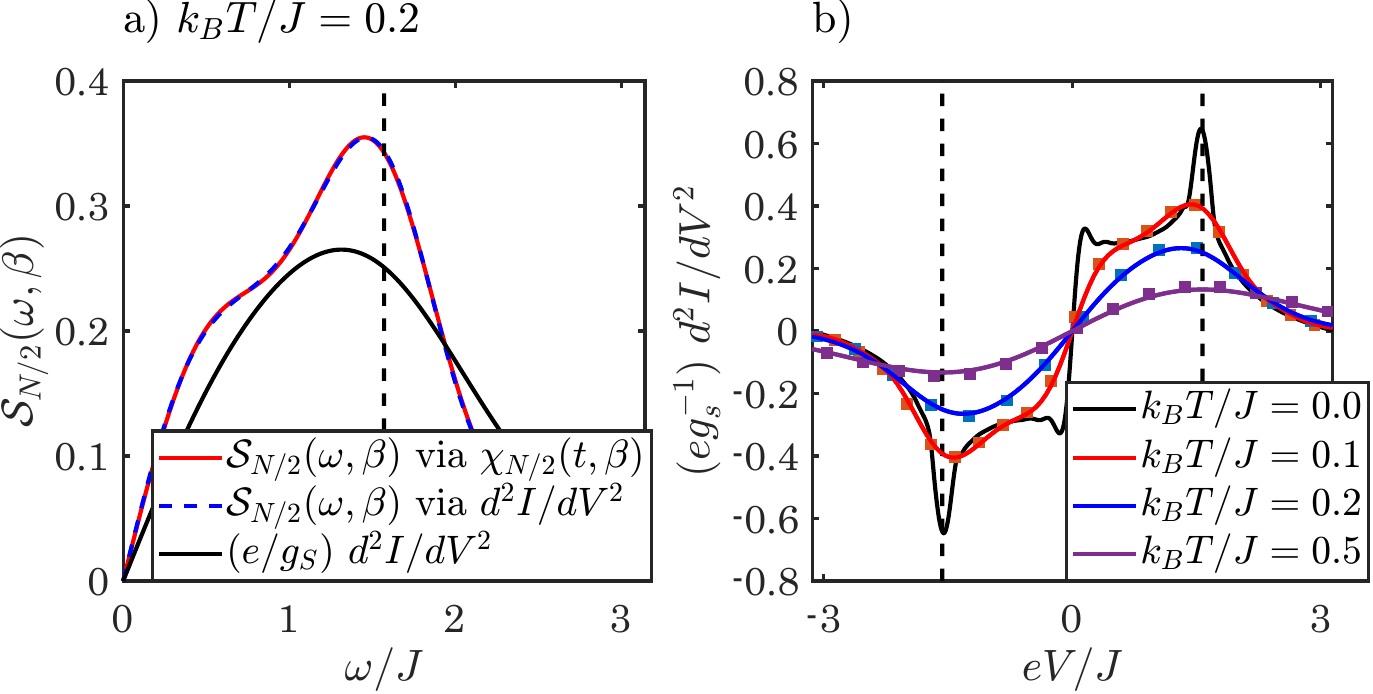}
\caption{\label{fig:thermal} 
The local spin spectral function, $\mathcal{S}_j(\omega, \beta)$, can be measured via the double differential conductance, $d^2I_j/dV^2$, in IETS. (a) Comparison of $\mathcal{S}_{j}(\omega, \beta)$ calculated directly from the local spin correlation function, $\chi_j(t,\beta)$, using \cref{eq:spectral_density} [red line] with $\mathcal{S}_{j}(\omega,\beta)$ extracted from the double differential conductance via \cref{eq:main_result} [dashed blue line]. The calculated $d^2I_j/dV^2$ is also shown [black line]. Calculations are for the spin-1/2 Heisenberg model, \cref{eq:heisenberg}, in the bulk (site $j=L/2$ in the spin chain) at $k_BT = 0.2J$. 
(b) Thermal broadening in $d^2I_j/dV^2$ [lines] for the spin-1/2 Heisenberg model matches the expected IETS thermal broadening obtained by convoluting $\mathcal{S}_{j}(\omega,\beta)$ with a Gaussian that has a half-width of $5.4k_BT$ [squares]~\cite{Jacklevic_lambe_original_inelastic_theory}. Both $\mathcal{S}_j(\omega, \beta)$ and $d^2I_j/dV^2$ show a Van Hove singularity at the des Cloizeaux-Pearson spin-wave frequency ($eV=\pi J/2$; dashed black line).}
\end{centering}
\end{figure}

\textit{Interpreting the (double) differential conductance of large magnetic systems.}
We plot the LDOS calculated from $d^2I/dV^2$, \cref{eq:DOS_relation}, for the spin-1/2 and spin-1 Heisenberg models in Figs. \ref{fig:spinhalf}a and \ref{fig:spinone}a, respectively. Except at $\omega=0$, see below, both agree with the LDOS calculated directly from the local dynamical spin correlators, \cref{eq:spin_dos_def}. This confirms that the LDOS and local spin spectral functions can be directly measured from the double differential conductance.

By relating the double differential conductance to the LDOS, distinct conductance characteristics for magnetic systems can be identified. For example, the bulk LDOS for the spin-$1/2$ Heisenberg chain, \cref{fig:spinhalf}a, displays a Van Hove singularity at the des Cloizeaux-Pearson spin-wave frequency of $eV=\pi J/2$ \cite{Cloizeaux_1962,Mueller_heisenberg_AFM}. This peak corresponds to the maximum in the lower bound of the spin-wave continuum and reveals itself as a peak in $d^2I/dV^2$, \cref{fig:spinhalf}b, or a 
kink in $dI/dV$, \cref{fig:spinhalf}c. We find that this conductance peak is present for all sites at $eV = \pi J / 2$, except for the edge sites. Here, the peak shifts to $eV\sim\pi J /4$ as the edge spin has only one nearest neighbor.

There are two Van Hove singularities in the bulk LDOS of the spin-1 Heisenberg chain, \cref{fig:spinone}a. These are located at $eV =0.41J$ and $eV=2.725J$, and correspond to the Haldane gap and the maximum in the single-magnon dispersion, respectively \cite{White_spectral_function}. Corresponding peaks (kinks) are found in the double (single) differential conductance, \cref{fig:spinone}b,c. The Haldane gap is clearly present in the (double) differential conductance at all sites. However, as we approach the edge of the chain, both Van Hove singularities broaden and the peak caused by the maximum in the single-magnon dispersion eventually vanishes, \cref{fig:spinone}b.


In \cref{fig:thermal}a, we compare the local spin spectral function calculated directly from the finite-temperature dynamical spin correlators, \cref{eq:spectral_density}, to the local spin spectral function from the double differential conductance, \cref{eq:main_result}. They are in excellent agreement. 

The thermal broadening of a single excitation in IETS can be approximated by a Gaussian with a half width of $5.4 k_BT$~\cite{Jacklevic_lambe_original,Jacklevic_lambe_original_inelastic_theory,Rossier_delago_review}. Thus, convoluting the spin spectral function, $\mathcal{S}_j(\omega,\beta)$, with a Gaussian should match the finite-temperature current, \cref{eq:double_diff}. We find that the tunneling current obtained by convolution and directly from the dynamical spin correlators are in excellent agreement, \cref{fig:thermal}b. 


The peak in the double differential conductance due to the Van Hove singularity for the spin-$1/2$ Heisenberg chain  remains visible to temperatures well above $k_BT=J/10$, \cref{fig:thermal}b. This is an eminently achievable temperature scale, even in systems with weak exchange coupling. For example, for $J = 10$~meV one has $J/10k_B\sim10K$.


\begin{figure}
\begin{centering}
\includegraphics[width=0.9\linewidth]{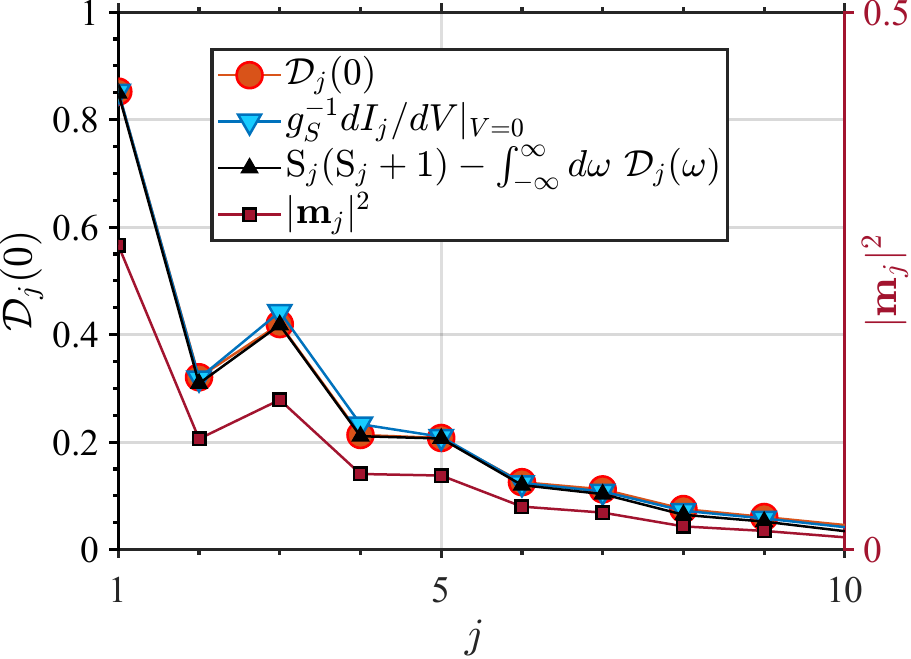}
\caption{\label{fig:zero_bias}  Zero-frequency modes in the spin-$1$ Heisenberg chain. The LDOS, $\mathcal{D}_j(0)$, can be 
(i) calculated directly, 
(ii) extracted from measurements of the zero-bias differential conductance, $dI_j/dV|_{eV=0}$, or 
(iii) from $d^2I_j / dV^2$ and the sum rule, $\mathrm{S}_j(\mathrm{S}_j + 1) = \int_{-\infty}^\infty d\omega\  \mathcal{D}_j(\omega)$, [see \cref{eq:sum_rule_dos}]. 
All three methods are in excellent agreement.
Approach (iii) works by determining $\mathcal{D}_j(\omega)$ from the double differential conductance via \cref{eq:DOS_relation}. This misses the zero-frequency modes due to the non-analyticity of the quantum fluctuation-dissipation theorem at $\omega=0$. Hence, the missing LDOS is ascribed to the zero-frequency mode.
$\mathcal{D}_j(0)$ exhibits the same exponential decay as the square of the local  magnetization, $|\textbf{m}_j|^2$ (right axis), as the chain is probed further away from the edge.}
\end{centering}
\end{figure}

\textit{Zero-frequency modes.} 
The local spin spectral functions and LDOS calculated directly from the dynamical spin correlator, $\chi_j(t,\beta)$, via \cref{eq:main_result,eq:spin_dos_def}, does not agree with the values determined from the double differential conductance, \cref{eq:DOS_relation}, at $\omega=0$. This is because Kubo's quantum fluctuation-dissipation theorem is non-analytic at $\omega = 0$~\cite{Kubo_1966}. Therefore, the zero-frequency modes are missed when $\mathcal{S}_j(0,\beta)$ and $\mathcal{D}_j(0)$ are extracted from the double differential conductance. For example, the zero-frequency peak in the LDOS near the edge site ($j=1$) for the Haldane phase is not captured when $\mathcal{D}_j(0)$ is calculated from $d^2I_j/dV^2$, \cref{fig:spinone}a.

One way to measure the zero-frequency LDOS is directly from the zero-bias differential conductance, $dI/dV$. Previously, it was established that the height of the differential conductance step corresponds to the local spin spectral weight at that voltage \cite{cyrus_single_atom_STM_IETS,heinrich_single_atom,Localprobes_heisenberg}. 
With the definition of the LDOS, \cref{eq:spin_dos_def}, this implies that 
\begin{equation}
\frac{dI}{dV}\bigg{|}_{V=0}=g_S\sum_{\substack{M_0,N_0,\\\alpha\in x,y,z}} |\langle M_0|\hat{S}_j^\alpha|N_0\rangle|^2 =g_S \mathcal{D}_j(0). \label{eq:zero_bias_diff_cond}
\end{equation}
As $\sum_\alpha |\langle N_0|S_j^\alpha|N_0\rangle|^2=|\textbf{m}_j|^2$, the zero bias differential current goes as the square of the local magnetization, $\textbf{m}_j=\langle( \hat{S}^x_j, \hat{S}^y_j, \hat{S}^z_j)\rangle$, \cref{fig:zero_bias}. Hence, in the spin-1 chain, ${dI}/{dV}{|}_{V=0}$ is large near the edge and vanishes in the bulk.




However, directly measuring the inelastic contribution to the zero-bias differential conductance can be experimentally difficult because additional current contributions can lead to zero-bias peaks~\cite{heinrich_single_atom, cyrus_single_atom_STM_IETS,Loth_heinrich_control_state_quantum_spins,Loth_heinrich_spin_spec,Haldane_nanographene,Ternes_2015,peter_2015}. For example, zero-bias Kondo peaks can appear near the boundaries of the Haldane chain from the spin-1/2 edge states, which are difficult to distinguish from the inelastic contribution~\cite{Mishra_nanographene, AKLT_long}.

Therefore, rather than directly measuring the zero-bias differential conductance, we propose that the zero-frequency modes of the LDOS can be calculated from the spectral sum rule associated with the total spin of the adatom, $\mathrm{S}_j(\mathrm{S}_j+1) = \int_{-\infty}^\infty d\omega\  \mathcal{S}_j(\omega,\beta)$. Because $\partial^2 I_j(V=0)/ \partial V^2 = 0$ [cf. \cref{fig:spinone}b], 
\begin{align}
\label{eq:sum_rule_spectral}
\mathcal{S}_j(0, \beta) & = \mathrm{S}_j(\mathrm{S}_j + 1)  \\
& \phantom{=} - \int_{-\infty}^\infty d \omega
\frac{1}{e g_S (1-e^{-\hbar\beta \omega})} \nonumber \\
& \phantom{=} \phantom{-}\times \int_{-\infty}^\infty \int_{-\infty}^\infty dt\  dV\frac{ e^{-i(\omega-eV)t} }{\mathcal{G}(t,\beta)}\frac{\partial^2 I_{j}(V)}{\partial V^2}, \nonumber
\end{align}
and
\begin{align}
\label{eq:sum_rule_dos}
\mathcal{D}_j(0) & = \mathrm{S}_j(\mathrm{S}_j + 1)  
- \int_{-\infty}^\infty d \omega
\frac{1}{e g_S (1-e^{-\hbar\beta \omega})} \frac{\partial^2 I_{j}(\omega)}{\partial V^2}. 
\end{align}
The zero-frequency LDOS arising from the spin-1/2 edge states in the spin-1 Heisenberg chain calculated via the sum over groundstates, \cref{eq:zero_bias_diff_cond},  is in excellent agreement with $\mathcal{D}_j(0)$ calculated from the double differential conductance and spectral sum rule via \cref{eq:sum_rule_dos}, \cref{fig:zero_bias}.

\label{sec:conclusion}
\textit{Conclusions.} We have shown that the local spin density of states can be measured from the double differential conductance of IETS experiments, for any system size. Relating the tunneling current to the spin LDOS allows us to interpret the  conductance features for spin chains in the thermodynamic limit. This shows that important spectral features often reveal themselves in the double differential conductance. Understanding how to interpret spectroscopic magnetic features in the absence of clearly distinguished, discrete steps in the differential conductance will play a key role in analyzing IETS experiments on large magnetic systems. By accessing the LDOS, zero-frequency modes can be measured from the spectral sum rules, allowing, e.g., the topological edge states of the Haldane phase to be probed. Detecting zero-frequency topological modes is also extremely important for other systems, such as Majorana topological edge modes in Kitaev spin chains \cite{Stephan_review_2025,Wiesendanger_majorana_2018}. 


\begin{acknowledgments}
We would like to thank Peter Jacobson for helpful conversations and a reading of the manuscript. This work was supported by the Australian Research Council (DP230100139) and the Australian Government Research
Training Program Scholarship.
\end{acknowledgments}

\bibliography{bibliography.bib}
%

\end{document}